\documentclass[aps,prl,superscriptaddress,showpacs,floatfix,onecolumn]{revtex4}
\usepackage{graphicx,amsmath,amssymb,dsfont,bm}

\usepackage{xcolor,hyperref}
\definecolor{dark-red}{rgb}{0.8,0.15,0.15}
\definecolor{dark-blue}{rgb}{0.15,0.15,0.4}
\definecolor{dark-green}{rgb}{0.1,0.5,0.1}
\definecolor{medium-blue}{rgb}{0,0,0.5}
\hypersetup{colorlinks, linkcolor={dark-red},citecolor={dark-green}, urlcolor={blue}}

\newcommand{\str}{\text{\,str\,}}
\newcommand{\tr}{\text{\,tr\,}}

\newcommand{\diag}{\text{diag\,}}

\begin{document}
\widetext

\title{Distribution of Scattering Matrix Elements in Quantum Chaotic Scattering}

\author{S. Kumar} \email{skumar.physics@gmail.com} 
\affiliation{Fakult\"at f\"ur Physik, Universit\"at Duisburg-Essen, Lotharstra\ss{}e 1, D-47048 Duisburg, Germany}
\author{A. Nock}
\affiliation{Fakult\"at f\"ur Physik, Universit\"at Duisburg-Essen, Lotharstra\ss{}e 1, D-47048 Duisburg, Germany}
\author{H.-J. Sommers}
\affiliation{Fakult\"at f\"ur Physik, Universit\"at Duisburg-Essen, Lotharstra\ss{}e 1, D-47048 Duisburg, Germany}
\author{T. Guhr} 
\affiliation{Fakult\"at f\"ur Physik, Universit\"at Duisburg-Essen, Lotharstra\ss{}e 1, D-47048 Duisburg, Germany}
\author{B. Dietz}
\affiliation{Institut f\"{u}r Kernphysik, Technische Universit\"{a}t Darmstadt, D-64289 Darmstadt, Germany}
\author{M. Miski-Oglu}
\affiliation{Institut f\"{u}r Kernphysik, Technische Universit\"{a}t Darmstadt, D-64289 Darmstadt, Germany}
\author{A. Richter}
\affiliation{Institut f\"{u}r Kernphysik, Technische Universit\"{a}t Darmstadt, D-64289 Darmstadt, Germany}
\author{F. Sch\"afer}
\affiliation{LENS, University of Florence, I-50019 Sesto-Fiorentino, Italy}

\begin{abstract}
Scattering is an important phenomenon which is observed in systems ranging from the micro- to macroscale. In the context of nuclear reaction theory the Heidelberg approach was proposed and later demonstrated to be applicable to  many chaotic scattering systems. To model the universal properties, stochasticity is introduced to the scattering matrix on the level of the Hamiltonian by using random matrices. A long-standing problem was the computation of the distribution of the off-diagonal scattering-matrix elements. We report here an exact solution to this problem and present analytical results for systems with preserved and with violated time-reversal invariance. Our derivation is based on a new variant of the supersymmetry method. We also validate our results with scattering data obtained from experiments with microwave billiards. 
\end{abstract}

\maketitle

\thispagestyle{empty}

\noindent
A large part of our knowledge about quantum systems comes from scattering experiments. Even in classical wave systems, observables can often be traced back to a scattering process~\cite{Scattering}. Important examples stem from nuclear, atomic and molecular physics, mesoscopic ballistic devices, and even from classical wave systems as, e.g., microwave and elastomechanical billiards, as well as from wireless communication~\cite{Scattering,BS1990,FKS2005,MRW2010,GGW1998,MW1969,MPS1985,MM2005,VWZ1985,Agassi1975,DB,LW1991,BLR1993,OT1993,F1996,FS1997,FSS2005,RFW,KMMS2005,KSW2005,H2005,Hul2005,L2008,D2009,D2010a,D2010b,EGLNO1996,AEOGS2010,Yeh2012,YOAA2012,BC2001}. Accordingly, the investigation of scattering phenomena has been a subject of major interest from both theoretical and experimental points of view. Here we focus on the universal features of chaotic scattering systems. 

The quantity of interest is the scattering matrix ($S$ matrix) which relates the asymptotic initial and final states, and owing to the flux conservation requirement is unitary. The $S$-matrix elements are given in terms of the Hamiltonian $H$ describing the scattering center~\cite{MW1969,FS1997} by
\begin{equation}
\label{Sab}
 S_{ab}(E)=\delta_{ab}-i 2\pi W_a^\dag G(E) W_b,
\end{equation}
where the inverse of the resolvent $G(E)$ reads
\begin{equation}
\label{GE}
G^{-1}(E)=E\mathds{1}_N-H+i\pi\displaystyle\sum_{c=1}^M W_c W_c^\dag.
\end{equation}
The coupling vectors $W_c$ account for the interaction between the internal states of $H$ and the $M$ open channel states labeled $c=1,...,M$ where the full system resides asymptotically before, respectively, after the scattering event. 
This ansatz yields the most general description of any scattering process in which an interaction zone and scattering channels can be identified. Without loss of generality we may restrict ourselves to a diagonal average $S$ matrix~\cite{EW1973}, that is, to orthogonal coupling vectors, viz., $W_c^\dag W_d=(\gamma_c/\pi) \delta_{cd}; c,d=1,\cdots ,M$ \cite{VWZ1985,FS1997}.

The chaoticity of the dynamics in the scattering center is taken into account by assuming that $H$ is a random matrix. This is referred to as the Heidelberg approach~\cite{MW1969} as distinguished from the Mexico approach~\cite{MPS1985,MM2005} where stochasticity is introduced in the $S$ matrix itself. In view of the universality conjecture~\cite{BGS}, the Hamiltonian $H$ is modeled by the Gaussian ensemble of $N\times N$ random matrices with the distribution~\cite{B1981,Mehta2004,GGW1998}, $\mathcal{P}(H)\propto \exp\big(-(\beta N/4v^2)\tr H^2\big)$. Here, $v^2$ fixes the energy scale and the index $\beta$ characterizes the symmetry class, i.e., the invariance properties of the Hamiltonian. We focus on the cases $\beta=1$ (Gaussian Orthogonal Ensemble, GOE) and $\beta =2$ (Gaussian Unitary Ensemble, GUE) which apply to systems with, respectively, preserved and violated time-reversal ($\mathcal{T}$) invariance~\cite{B1981,Mehta2004,GGW1998}; the former being rotationally invariant as well. The universal spectral properties of closed chaotic systems deduced from these ensembles are by now well understood~\cite{GGW1998}. Those of the effective Hamiltonian associated with a chaotic scattering process, which includes $H$ and the coupling to the exterior have been derived, e.g., in~\cite{FS1997}. Its complex eigenvalues can be extracted from the measured resonance spectra~\cite{Kuhl2008,MRW2010} in general only in the regimes of isolated and weakly overlapping resonances. Accordingly a description of the universal properties of the $S$ matrix itself is indispensable.

In their pioneering work~\cite{VWZ1985}, using the supersymmetry method~\cite{Efetov}, Verbaarschot {\it et al.} calculated the energy correlation function of two $S$-matrix elements. Further progress in this direction was made in~\cite{DB} where three- and four-point $S$-matrix correlation functions were evaluated. While these provide rich information about the scattering process, a full statistical description requires the determination of the distributions of the $S$-matrix elements. Their knowledge yields information about all their moments and is highly desirable also from an experimental viewpoint~\cite{MRW2010}. The distributions and higher moments have been known hitherto only in the limit of a large number $M$ of open channels and a vanishing average $S$ matrix, i.e., in the Ericson regime~\cite{MRW2010,Agassi1975} or in a high-loss environment~\cite{Yeh2012,YOAA2012}. There the real and the imaginary parts of the $S$-matrix elements are Gaussian distributed. Otherwise the deviations from this behavior are significant due to the unitarity of the $S$ matrix~\cite{MRW2010,DB,T1975,RSW1975}. The complexity involved in the calculations of the correlation functions~\cite{VWZ1985,DB,StockGuhr2004} indicates that those of the distributions of the $S$-matrix elements constitute a challenging task. However, this was partially accomplished in~\cite{FSS2005} where the distribution of the diagonal $S$-matrix elements was derived. Moreover, in ~\cite{RFW} the statistics of transmitted power, viz. $|G_{nm}(E)|^2, n\neq m$, was calculated. These results have been verified in microwave experiments~\cite{KMMS2005,KSW2005,H2005,Hul2005,L2008,D2009,D2010a,D2010b}. In the present Letter, we provide analytical results for the distributions of the off-diagonal $S$-matrix elements which could not be computed with the well-established methods~\cite{VWZ1985,DB,FSS2005}. The novelty of our approach lies in that a nonlinear sigma model is constructed based on the {\it characteristic function} associated with the distributions which is the generating function for the moments. In contrast, the standard supersymmetry approach starts from the generating function for the $S$-matrix correlations.

We introduce the notation $\wp_s(S_{ab})$, with $s=1,2$ to refer to the real and imaginary parts of $S_{ab}$, respectively. Thus Eq.~\eqref{Sab} yields for the off-diagonal ($a\neq b$) elements
\begin{eqnarray}
\wp_s(S_{ab})=\pi \big((-i)^sW_a^\dag G W_b +i^s W_b^\dag G^\dag W_a\big). 
\end{eqnarray}
Determining distributions for $\wp_s(S_{ab})$, which we denote by $P_s(x_s)$, involves the nontrivial task of performing an ensemble average,
\begin{equation}
\label{Ps1}
P_s(x_s)=\int d[H] \mathcal{P}(H)\delta(x_s-\wp_s(S_{ab})),~~~ s=1,2.
\end{equation}
We instead first compute the corresponding characteristic function,
\begin{equation}
R_s(k)=\int d[H]\mathcal{P}(H) \exp(-i k \wp_s(S_{ab})),
\end{equation}
and then obtain $P_s(x_s)$ as the Fourier transform of $R_s(k)$,
\begin{equation}
\label{PsRs}
P_s(x_s)=\frac{1}{2\pi}\int_{-\infty}^\infty dk  R_s(k) \exp(i k x_s).
\end{equation}
Defining the $2N$-component vector $W$ and the $2N\times 2N$ matrix $A_s$ as
\begin{equation}
W=\begin{bmatrix}W_a \\ W_b \end{bmatrix}, ~~~A_s=\begin{bmatrix} 0 & (-i)^s G \\ i^s G^\dag & 0 \end{bmatrix}, 
\end{equation}
we rewrite $R_s(k)$ as
\begin{equation}
R_s(k)=\int d[H] \mathcal{P}(H) \exp(-i k \pi W^\dag A_s W).
\end{equation}
In this form, the ensemble average can not be performed, because $A_s$ contains the inverse of $H$. To carry it out, we map the statistical model to superspace. We introduce the $2N$-vectors $z^T=[z_a^T, z_b^T]$ and  $\zeta^T=[\zeta_a^T, \zeta_b^T]$ consisting of complex commuting and anticommuting (Grassmann) variables, respectively. The supervector is constructed in the usual manner~\cite{Guhr2010} as $\Psi^T=[z^T,\zeta^T]$.
Using these vectors, and multivariate Gaussian-integral results, we recast the characteristic function as
\begin{eqnarray}
 R_s(k)=\int d[\Psi] \exp\Big(\frac{i}{2}(\mathbf{W}^\dag \Psi+\Psi^\dag \mathbf{W})\Big)\int d[H] \mathcal{P}(H)\exp\Big(\frac{i}{4\pi k} \Psi^\dag \mathbf{A}_s^{-1} \Psi\Big).
\end{eqnarray}
Here $\mathbf{A}_s^{-1}=\mathds{1}_2\otimes A_s^{-1}$, and $\mathbf{W}^\dag=[W^\dag,0]$ is a $4N$-vector. The ensemble average can now be done, leading to an enormous reduction in the {\it degrees of freedom}. To facilitate it we block diagonalize $\mathbf{A}_s^{-1}$ by the transformations
\begin{eqnarray}
\label{transform}
\nonumber
&z\rightarrow T^{+} z,\, z^\dag \rightarrow z^\dag;
~~\zeta\rightarrow \frac{2}{\beta}T^{-}\,\zeta,\, \zeta^\dag \rightarrow \zeta^\dag,\\
&T^{\pm}=\begin{bmatrix} 0 & \pm(-i)^s\mathds{1}_N \\ -i^s\mathds{1}_N & 0 \end{bmatrix}.
\end{eqnarray}
Here, we have used the fact that the complex quantities $z\,\, (\zeta)$ and $z^\dag\, (\zeta^\dag)$ are independent of each other. The Jacobian of the transformation is $(-1)^N 2^{2(\beta-2)N}$. After the application of Eq.~(\ref{transform}) we have to distinguish between the two cases.

For $\beta=2$ we obtain
\begin{eqnarray}
\label{RkUE1}
 R_s(k)=(-1)^N\int d[\Psi] \exp\Big(\frac{i}{2}(\mathbf{U}_s^\dag \Psi+\Psi^\dag \mathbf{W})\Big)\int d[H] 
\mathcal{P}(H)\exp\Big(\frac{i}{4\pi k} \Psi^\dag \boldsymbol{\mathcal{A}}^{-1} \Psi\Big),
\end{eqnarray}
with the $4N$-vector $\mathbf{U}_s^\dag=\big[-i^sW_b^\dag,(-i)^sW_a^\dag,0,0\big]$ and matrix $\boldsymbol{\mathcal{A}}^{-1}=\text{diag}\left[-(G^{-1})^\dag, G^{-1},-(G^{-1})^\dag,-G^{-1}\right]$, and $\Psi^T=[z^T,\zeta^T]$ as above. 

For $\beta=1$ we decompose the $2N$-vector $z$ into its real and imaginary parts $x$ and $y$ (not to be confused with $x_1$ and $x_2$) to construct a $4N$-vector. In addition, we symmetrize the vector $\zeta$ using $\zeta_a^*, \zeta_b^*$ along with $\zeta_a, \zeta_b$, thereby doubling its size as well. The associated Jacobian equals $2^{2N}$ and thus cancels that of the transformation Eq.~(\ref{transform}). The $8N$-supervector is given as $\Psi^\dag=[x_a^T,y_a^T,x_b^T,y_b^T,\zeta_a^\dag,-\zeta_a^T,\zeta_b^\dag,-\zeta_b^T]$ and 
\begin{eqnarray}
\label{RkOE1}
 R_s(k)=(-1)^N\int d[\Psi] \exp\big(i\Psi^\dag \mathbf{V}_s\big)\int d[H] 
\mathcal{P}(H)\exp\Big(\frac{i}{4\pi k} \Psi^\dag \boldsymbol{\mathcal{A}}^{-1} \Psi\Big).
\end{eqnarray}
In this case $\mathbf{V}_s^T=\left[i^s w_s^{-},-i^{s+1}w_s^{+},w_s^{+},iw_s^{-},0,0,0,0\right]$, where $w_s^{\pm}=((-i)^s\,W_a^T\pm W_b^T)/2$, and $\boldsymbol{\mathcal{A}}^{-1}=\text{diag}[-(G^{-1})^\dag,G^{-1},-(G^{-1})^\dag,-G^{-1}]\otimes\mathds{1}_{2}$. Since the matrix $\boldsymbol{\mathcal{A}}^{-1}$ is block diagonal in both Eqs. \eqref{RkUE1}, \eqref{RkOE1} the ensemble averaging is now straightforward.

Next we use the Hubbard-Stratonovitch identity~\cite{Efetov,Guhr2010}  to map the integral over the $8N/\beta$-supervector $\Psi$ to a matrix integral in superspace involving an $8/\beta$-dimensional supermatrix $\sigma$ of appropriate symmetry. This yields 
\begin{eqnarray}
\label{Rk_sigma}
R_s(k)= \int d[\sigma] \exp\Big(-r\str\sigma^2-\frac{\beta}{2}\str\ln\boldsymbol{\Sigma}-\frac{i}{4}\mathbf{F}_s\Big),~~\\
\nonumber
\boldsymbol{\Sigma}=\sigma_E\otimes \mathds{1}_N+\frac{i} {4k}L\otimes \sum_{c=1}^M W_c W_c^\dag,~\sigma_E=\sigma-\frac{E}{4\pi k}\mathds{1}_{8/\beta},
\end{eqnarray}
with str denoting the supertrace. Here $r=(4\beta\pi^2k^2 N)/v^2$, and $L =\diag(1,-1,1,-1)\otimes \mathds{1}_{2/\beta}$.  $\mathbf{F}_s$ equals $\mathbf{V}_s^T \mathbf{L}^{-1/2}\boldsymbol{\Sigma}^{-1}\mathbf{L}^{-1/2}\mathbf{V}_s$ for $\beta=1$, and $\mathbf{U}_s^\dag \mathbf{L}^{-1/2}\boldsymbol{\Sigma}^{-1}\mathbf{L}^{-1/2}\mathbf{W}$ for $\beta=2$ with $\mathbf{L}=L\otimes \mathds{1}_N$. The supersymmetric representation, Eq. ~\eqref{Rk_sigma}, constitutes one of our key results.

The orthogonality of $W_c$ leads to
\begin{eqnarray}
\label{expansions}
\nonumber
\str \ln \boldsymbol{\Sigma}=N \str\ln \sigma_E+\sum_{c=1}^M \str\ln\Big(\mathds{1}_{8/\beta}+\frac{i\gamma_c}{4\pi k} \sigma_E^{-1}L \Big),~~~~~\\
\boldsymbol{\Sigma}^{-1}=\sigma_E^{-1}\otimes\mathds{1}_N-\sigma_E^{-1}\otimes\sum_{c=1}^M \frac{\pi}{\gamma_c}W_c W_c^\dag+\sum_{c=1}^M \rho^{(c)}\otimes \frac{\pi}{\gamma_c}W_c W_c^\dag,
\end{eqnarray}
with $\rho^{(c)}=(\sigma_E+ i\gamma_c/(4\pi k)L)^{-1}$. Furthermore, $\mathbf{F}_s$ equals a linear combination of matrix elements of $\rho^{(c)}$ multiplied with $\gamma_c$, where $c=a,\, b$. In Eq.~\eqref{expansions} the first term is of order $N$ while the rest is of order $1$. Thus, in order to perform the limit $N\rightarrow \infty$, we may apply the saddle point approximation. This leads to a separation of $\sigma$ into Goldstone modes $\sigma _G$ and massive modes~\cite{Guhr2010}. The integrals over the latter, being Gaussian ones, can be readily done and yield unity. We are therefore left with an expression involving only the Goldstone modes, and consequently our sigma model reads
\begin{equation}
\label{RkG}
R_s(k)=\int d\mu(\sigma_G) \exp\Big({-\frac{i}{4}\mathbf{F}_s}\Big) \prod_{c=1}^M\text{sdet}^{-\beta/2}\Big(\mathds{1}_{8/\beta}+\frac{i\gamma_c}{4\pi k} \sigma_E^{-1}L\Big),
\end{equation}
with sdet denoting the superdeterminant and $\sigma$ replaced by $\sigma_G$ in all the ingredients of Eq.~(\ref{Rk_sigma}). In order to perform the remaining integrals we proceed as in~\cite{VWZ1985,Guhr2010} and express $\sigma_G$ in terms of an $8/\beta$-dimensional supermatrix $Q$ as $\sigma_G=(E/8\pi k)\mathds{1}_{8/\beta}-(\Delta/8\pi k)Q$ with $Q^2=-\mathds{1}_{8/\beta}$. Here, $\Delta=(4v^2-E^2)^{1/2}$ with $\Delta/(2\pi v^2)$ identified as the celebrated Wigner semicircle.
We use the parametrization of $Q$ as in \cite{FS1997,VWZ1985,Verbaarschot1988}. For $\beta=2$, it involves pseudo eigenvalues $\lambda_1\in(1,\infty),\lambda_2\in(-1,1)$, angles $\phi_1,\phi_2\in(0,2\pi)$ and four Grassmann variables. For $\beta=1$ we have three pseudo eigenvalues $\lambda_0\in(-1,1),\lambda_1,\lambda_2\in(1,\infty)$, two O(2) angles $\phi_1,\phi_2\in(0,2\pi)$, three SU(2) variables $m,r,s\in(-\infty,\infty)$, and eight Grassmann variables. The product over the superdeterminants in Eq.~\eqref{RkG} involves the pseudo eigenvalues only, viz.,
\begin{eqnarray*}
\begin{matrix}
 \mathcal{F}_\text{O}=\displaystyle \prod_{c=1}^M \dfrac{g_c^{+}+\lambda_0}{(g_c^{+}+\lambda_1)^{1/2}(g_c^{+}+\lambda_2)^{1/2}}  & \text{ for } \beta=1,\\
 \mathcal{F}_\text{U}=\displaystyle\prod_{c=1}^M \dfrac{g_c^{+}+\lambda_2}{g_c^{+}+\lambda_1}~~~~~~~~~~~~~~~~~~~~~~& \text{ for } \beta=2.
 \end{matrix}
\end{eqnarray*}
Here $g_c^{\pm}=(v^2\pm\ \gamma_c^2)/(\gamma_c \Delta)$. $g_c^{+}$ is related to the transmission coefficient $T_c=1-|\overline{S_{cc}}|^2$ as $g_c^{+}=2/T_c-1$~\cite{VWZ1985}. The exponential part in Eq.~\eqref{RkG} also involves other variables and is quite complicated for $\beta=1$. 

For $\beta=2$ the integrals over the Grassmann variables and the angles can be performed and we obtain the same distribution for the real and imaginary parts,
\begin{eqnarray}
\label{RkUE}
R_s(k)=1-\int_1^\infty d\lambda_1\int_{-1}^1 d\lambda_2 \frac{k^2}{4(\lambda_1-\lambda_2)^2}\,\mathcal{F}_\text{U}(\lambda_1,\lambda_2) \big(t_a^1 t_b^1+t_a^2 t_b^2\big)J_0\Big(k \sqrt{t_a^1 t_b^1}\Big),~~~~
\end{eqnarray}
where $J_n(z)$ represents the $n$th order Bessel function of the first kind, and
$t_c^j=\sqrt{|\lambda_j^2-1|}/(g_c^{+}+\lambda_j),\, j=1,2$. The ``1" in Eq.~\eqref{RkUE} is an Efetov-Wegner contribution~\cite{Efetov} which is essential for the correct normalization, $R_s(0)=1$. The distribution function is obtained using Eq. ~\eqref{PsRs} as
\begin{eqnarray}
\label{PxUE}
\nonumber
 P_s(x_s)=\frac{\partial^2 f(x_s)}{\partial x_s^2}\,;\hspace{4cm}\\
 f(x)=x\Theta(x)+\int_1^\infty d\lambda_1\int_{-1}^1 d\lambda_2\, \frac{\mathcal{F}_\text{U}(\lambda_1,\lambda_2)}{4\pi(\lambda_1-\lambda_2)^2}\big(t_a^1 t_b^1+t_a^2 t_b^2\big)\big(t_a^1 t_b^1-x^2\big)^{-1/2}\Theta(t_a^1 t_b^1-x^2).~
\end{eqnarray}
Here, $\Theta(u)$ is the Heaviside function. The distributions being identical for $s=1,2$ in this case, the phases have a uniform distribution and the joint density of the real and the imaginary parts depends on $\sqrt{x_1^2+x_2^2}$ only. This facilitates the calculation of the distribution of their moduli~\cite{Details} and those of the cross sections which are given by the squared-moduli~\cite{MRW2010}. This is of particular relevance for the experiments where only these are accessible.

For $\beta=1$, the calculation involved is rather cumbersome. Nevertheless, we managed to perform all but four integrals. We have
\begin{eqnarray}
\label{RkOE}
R_1(k)=1+\frac{1}{8\pi}\int_{-1}^1 \!d\lambda_0 \int_1^\infty \!d\lambda_1\int_1^\infty \!d\lambda_2 \int_0^{2\pi}\!d\psi 
\mathcal{J}(\lambda_0,\lambda_1,\lambda_2)\mathcal{F}_\text{O}(\lambda_0,\lambda_1,\lambda_2)\sum_{n=1}^4\kappa_n k^n,~~
\end{eqnarray}
where
\begin{eqnarray*}
\mathcal{J}=\frac{(1-\lambda_0^2)|\lambda_1-\lambda_2|}
{2(\lambda_1^2-1)^{1/2}(\lambda_2^2-1)^{1/2}(\lambda_1-\lambda_0)^2(\lambda_2-\lambda_0)^2}.
\end{eqnarray*}
The $\kappa$'s entering Eq. (\ref{RkOE}) are functions of
\begin{eqnarray}
\label{qcpm}
\nonumber
p_c^j=\frac{\sqrt{|\lambda_j^2-1|}}{8(g_c^{+} +\lambda_j)},~ j=0,1,2;~~
p_c^{\pm}=p_c^1\pm p_c^2,~~~\\
\nonumber
q_c^{+}=\frac{1}{8}\Big(\frac{E}{\Delta}+i g_c^{-}\Big)\Big(\frac{1}{g_c^{+}+\lambda_1}+\frac{1}{g_c^{+}+\lambda_2}
-\frac{2}{g_c^{+}+\lambda_0}\Big),\\
q_c^{-}=\frac{1}{8}\Big(\frac{E}{\Delta}+i g_c^{-}\Big)\Big(\frac{1}{g_c^{+}+\lambda_1}-\frac{1}{g_c^{+}+\lambda_2}\Big),~~~~~~~~~~
\end{eqnarray}
and of the complex conjugate of $q_c^{\pm}$,
 $r_c^{\pm}=(q_c^{\pm})^*$,  and the quantities $\omega=2\sqrt{X Y},~~ l=X/Y, ~~m=Y/X,$
where
$
X=2p_a^{+}+q_a^{-}e^{-i2\psi}+r_a^{-}e^{i2\psi}$ and
$Y=2p_b^{+}-q_b^{-}e^{i2\psi}-r_b^{-}e^{-i2\psi}.
$
It can be verified that $\omega^2$ is real and takes values from the interval $[0,1]$. The $\kappa$'s are given as
\begin{eqnarray*}
\nonumber
&& \kappa_1=\kappa_{11} J_1(k\omega),~~\kappa_2=\kappa_{21} J_0(k \omega)+\kappa_{22} J_2(k \omega),\\
\nonumber
 &&\kappa_3=\kappa_{31} J_1(k\omega)+\kappa_{32} J_3(k\omega),\\
 &&\kappa_4=\kappa_{41} J_0(k\omega)+\kappa_{42} J_2(k\omega)+\kappa_{43} J_4(k\omega),
\end{eqnarray*}
with the entries $\kappa_{ij}$ given in the Appendix. $R_2(k)$ is obtained by multiplying $-i$ to the right-hand side of the expressions for $q_c^{\pm}$ in Eq.~\eqref{qcpm}, and changing $r_c^{\pm}$ accordingly. The distribution is obtained as
\begin{equation*}
P_s(x_s)=\delta(x_s)+\frac{\partial f_1}{\partial x_s}+\frac{\partial^2 f_2}{\partial x_s^2}+\frac{\partial^3 f_3}{\partial x_s^3}+\frac{\partial^4 f_4}{\partial x_s^4};~~~~~~~~
\end{equation*}
\begin{eqnarray}
\label{Px}
\nonumber
&&f_1=\left\langle\kappa_{11} x_s/\omega
\right\rangle,
~~f_2=-\left\langle\kappa_{21}+\kappa_{22}\big(1-2x_s^2/\omega^2\big)
\right\rangle,\\
\nonumber
&&f_3=-\left\langle\big[\kappa_{31}+\kappa_{32}\big(3-4x_s^2/\omega^2\big)\big]x_s/\omega\right\rangle,\\
&&f_4=\big\langle\kappa_{41}+\kappa_{42}\big(1-2x_s^2/\omega^2\big)
+\kappa_{43}\big(1-8x_s^2/\omega^2+8x_s^4/\omega^4\big)
\big\rangle.
\end{eqnarray}
Here the angular brackets represent the following:
\begin{eqnarray*}
\langle h \rangle=\frac{1}{16\pi^2}\int_{-1}^1 \!d\lambda_0 \int_1^\infty \!d\lambda_1\int_1^\infty \!d\lambda_2 \int_0^{2\pi}\!d\psi \mathcal{J}(\lambda_0,\lambda_1,\lambda_2)
\mathcal{F}_\text{O}(\lambda_0,\lambda_1,\lambda_2)\, 2h (\omega^2-x_s^2)^{-1/2}\Theta(\omega^2-x_s^2).
\end{eqnarray*}
Different results for the real and imaginary parts explain their unequal deviations from a Gaussian behavior which was observed in~\cite{T1975,RSW1975}. Details of the supersymmetry calculations and further results are given elsewhere~\cite{Details}.\\

\begin{figure}[ht]
\includegraphics[width=0.52\textwidth]{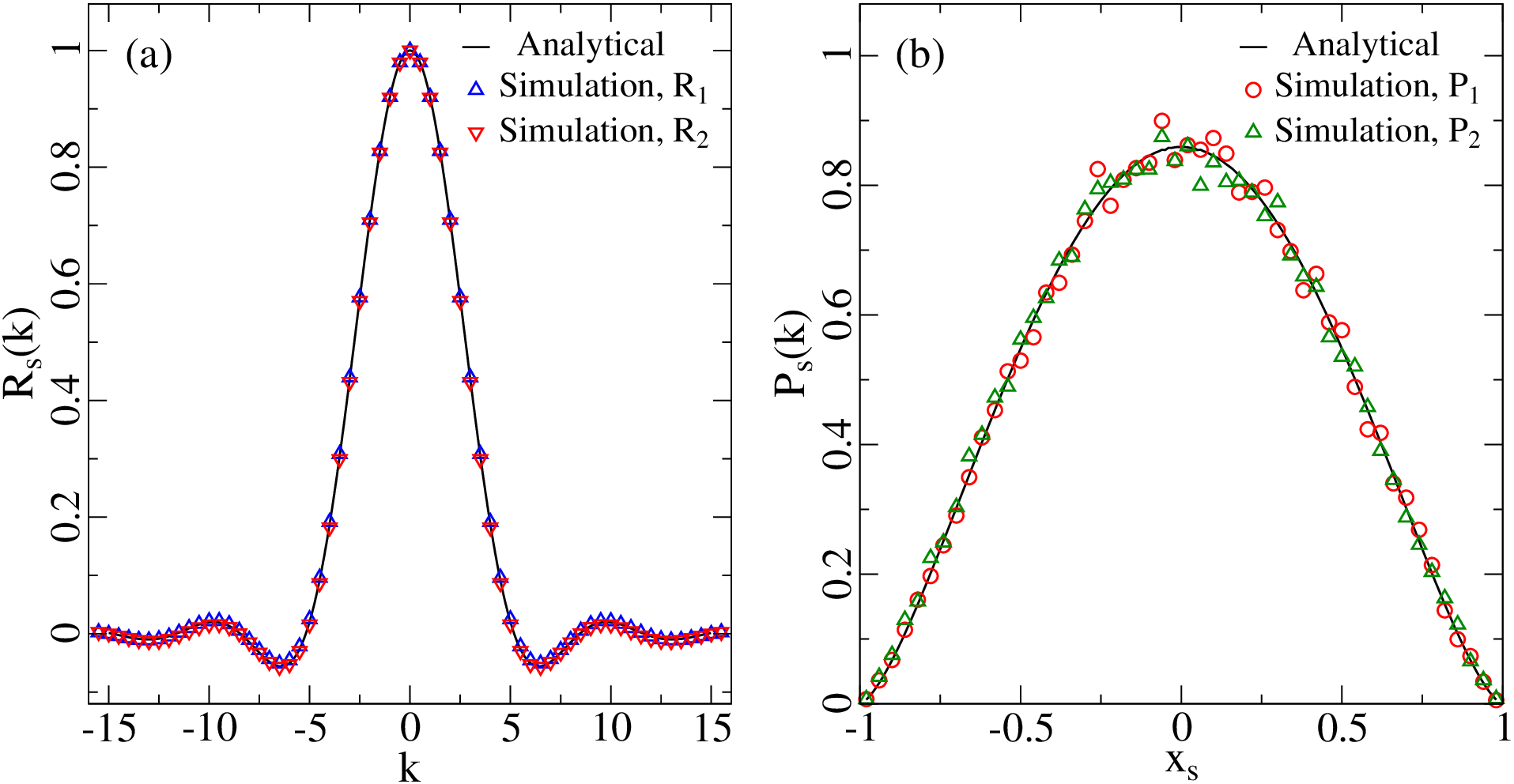}
\caption{Comparison of analytical and simulation results for $\beta=2$: (a) Characteristic functions (b) Distributions of the real and the imaginary parts of $S_{ab}$ for the choice of parameters $M=3, E=0.25, v=1, \gamma_1=0.8, \gamma_2=1.0,\gamma_3=1.2, a=1,b=2$.}
\label{fig1}
\end{figure}

We evaluated Eqs.~(\ref{RkUE}) and~(\ref{PxUE}) numerically using {\sc mathematica}~\cite{Mathematica}. The corresponding  {\sc mathematica} codes are included in the supplemental material~\cite{Supplemental}. In Fig. \ref{fig1} we compare for $\beta=2$ the analytical results for characteristic functions and distributions with simulations obtained with an ensemble of 50000 random matrices $H$ of dimensions $200\times200$ from the GUE~\cite{GGW1998,Mehta2004}. The agreement is excellent. Unfortunately, there were no experimental data available for this case because a complete $\mathcal{T}$ invariance violation could not be achieved. 
For $\beta=1$ we found that $R_s(k)$ is best evaluated using the Efetov variables $\theta_0,\theta_1,\theta_2$  ($0<\theta_0<\pi, 0<\theta_{1,2}<\infty$)~\cite{Efetov}. These are related to the $\lambda$'s as $\lambda_0=\cos\theta_0~$ and $\lambda_{1,2}=\cosh(\theta_1\pm\theta_2)$. The numerical evaluation of the fourth derivative needed for the computation of $P_s(x_s)$ is not feasible. We therefore instead determined them with the help of Eq.~(\ref{PsRs}), considering a cut-off for $k$. This approach works well for a sufficiently flat distribution, whereas, if it is highly localized, it is advantageous to consider the corresponding characteristic function instead. We found that the analytical results converge to the expected Gaussian distributions in the Ericson regime for both $\beta$ values~\cite{Details}.

\begin{figure}[ht]
\includegraphics[width=0.52\textwidth]{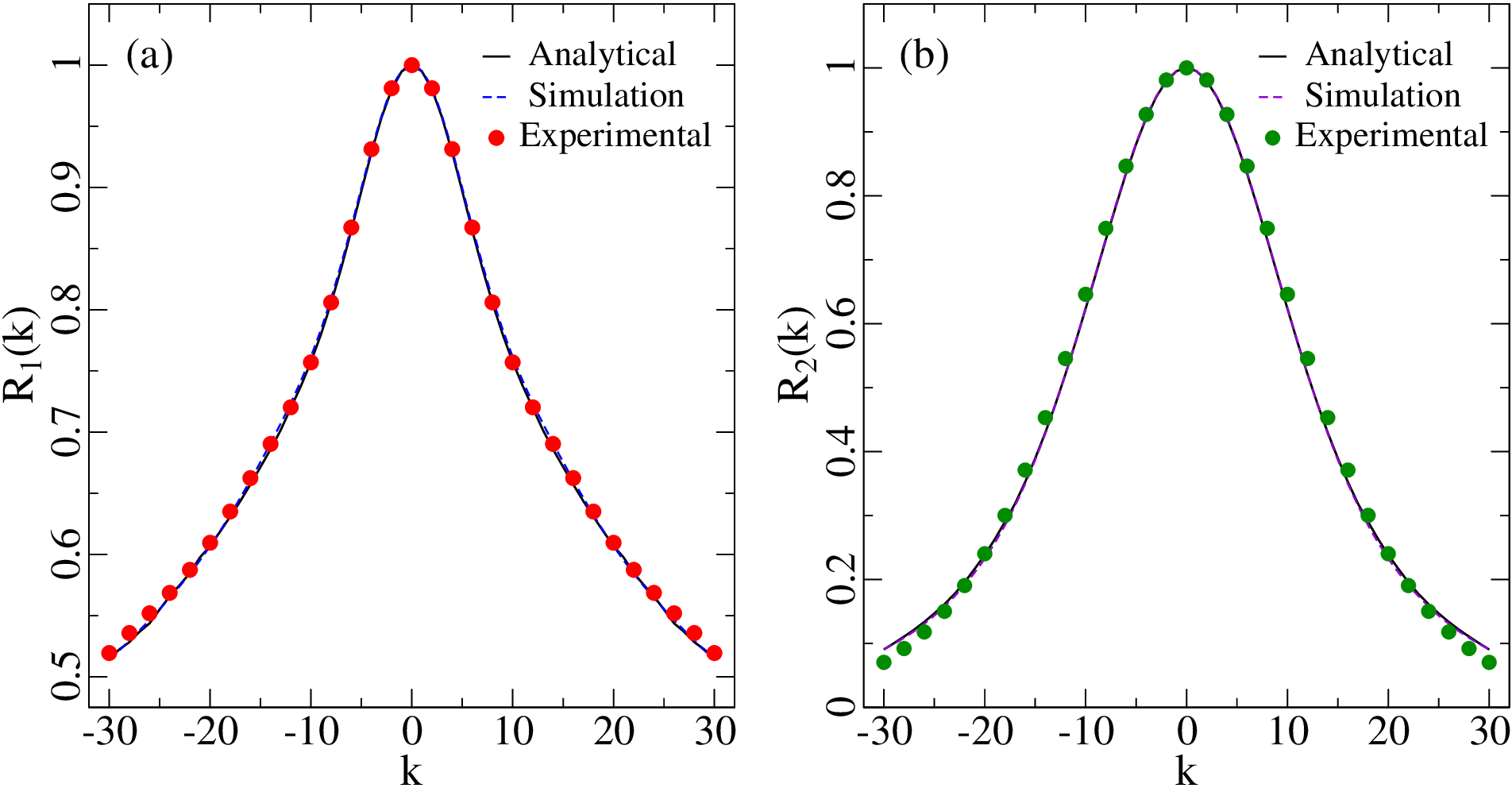}
\caption{Characteristic functions $R_1$ and $R_2$ corresponding to the real and the imaginary parts of $S_{12}$ for $\beta=1$. Comparison between the analytical results and the microwave experiment data for the frequency range 10-11 GHz~\cite{D2009,D2010a,D2010b}.} 
\label{fig2}
\end{figure}

\begin{figure}[ht]
\includegraphics[width=0.52\textwidth]{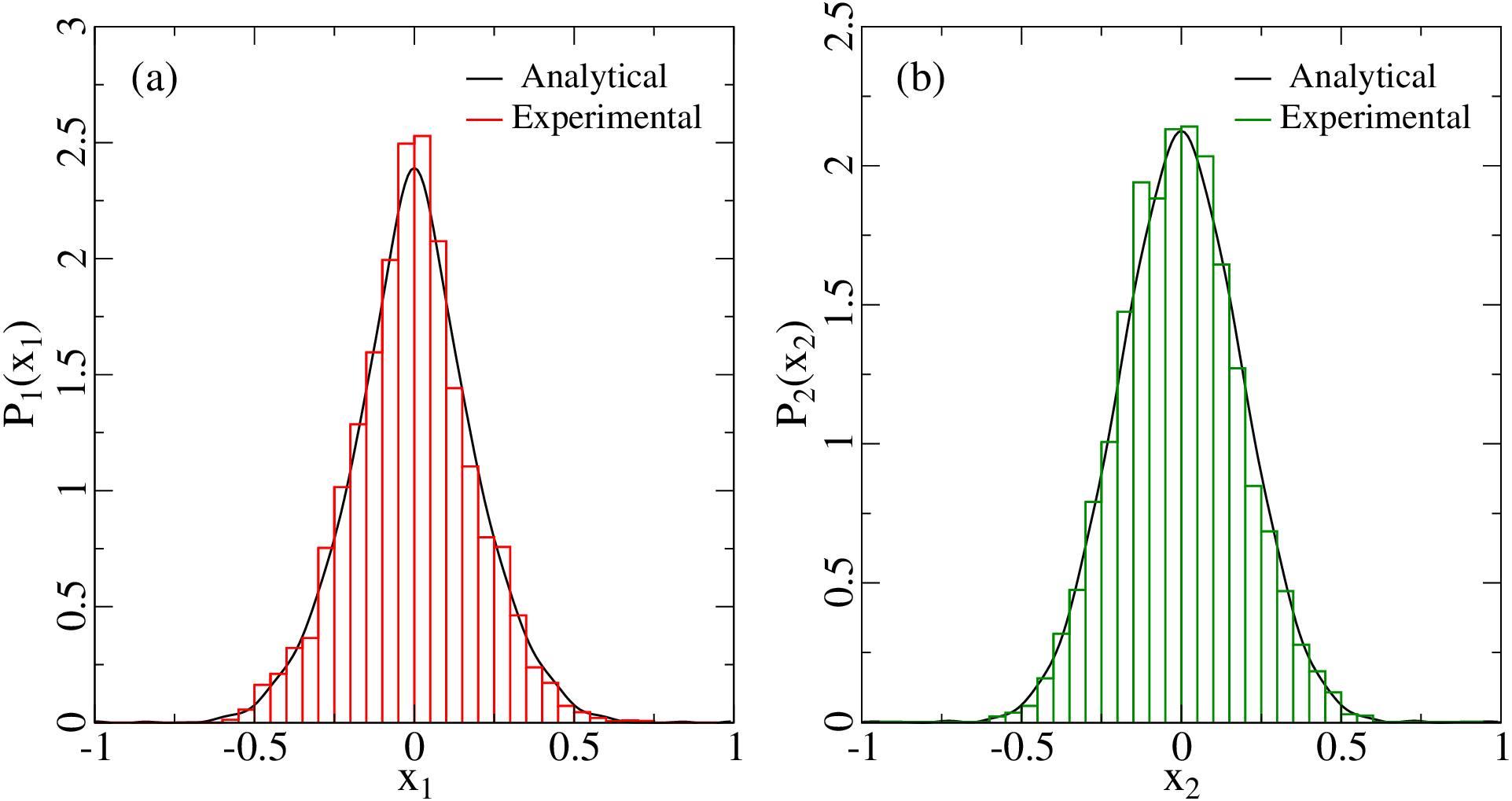}
\caption{Distributions of the real ($P_1$) and the imaginary ($P_2$) parts of $S_{12}$ for $\beta=1$. Comparison between analytical results and microwave experiment data for the frequency range 24-25 GHz~\cite{D2009,D2010a,D2010b}.}
\label{fig3}
\end{figure}

We tested our analytical results for $\beta =1$ with experimental data. To realize a chaotic scattering system, a microwave billiard with the shape of a classically chaotic tilted-stadium~\cite{D2009,D2010a,D2010b,PS1994} billiard was chosen and the resonator modes were coupled to the exterior via two antennas attached to it. An ensemble of several chaotic systems was obtained by introducing a small scatterer into the microwave billiard and moving it to six different positions~\cite{Mendez2003}. For the determination of the $S$-matrix elements a vector network analyzer coupled microwave power into and out of the resonator via the antennas. The frequency range was chosen such that only the vertical component of the electric field strength was excited. Then the Helmholtz equation is mathematically equivalent to the Schr\"odinger equation of the quantum tilted stadium billiard. The $S$-matrix elements were measured in steps of 100~kHz in a range from 1-25 GHz and the fluctuation properties of the $S$-matrix elements were evaluated in frequency windows of 1~GHz in order to guarantee a negligible secular variation of the coupling vectors $W_c$. More details concerning the experimental setup and the measurements are provided in~\cite{D2009,D2010a}. In Figs. \ref{fig2} and \ref{fig3}, we test the analytical results with experimental data for the frequency ranges 10-11~GHz and 24-25~GHz, corresponding to a ratio of the average resonance width $\Gamma$ and average resonance spacing $d$, $\Gamma/d=0.234$ and, respectively, $\Gamma/d=1.21$. The agreement is very good.

To conclude, we solved the long-standing problem of deriving the full distribution of the off-diagonal $S$-matrix elements valid in all regimes. We accomplished this task by introducing a novel route to the sigma model based on the characteristic function. We verified our analytical results with numerical simulations and with experimental data and found excellent agreements, and thus presented a new confirmation of the random matrix universality conjecture. 

\acknowledgments

This work was supported by the Deutsche Forschungsgemeinschaft within the Collaborative Research Centers SFB/TR12, ``Symmetries and Universality in Mesoscopic Systems'' and SFB634 (subproject ``Quantum Chaos and Wave-Dynamical Chaos''). 


\section{Appendix: Explicit expressions for the $\kappa_{ij}$}
\begin{equation*}
\kappa_{11}=-(9/8)\{p_a^{+}m^{1/2} \}_{+},
\end{equation*}
\begin{eqnarray*}
\kappa_{21}=-(1/4)(128 p_a^0p_b^0+14p_a^{+}p_b^{+}+32p_a^{-}p_b^{-})
+\{3e^{i2\psi}(p_a^{-}q_b^{+}-p_b^{-}r_a^{+})\}_{-}
+\{e^{-4i\psi}q_a^{-}r_b^{-}\}_{+},
\end{eqnarray*}
\begin{equation*}
\kappa_{22}=-(1/4)\{(p_a^{+}p_a^{+}-4q_a^{-}r_a^{-})m\}_{+}, 
\end{equation*}
\begin{eqnarray*}
\nonumber
&&\kappa_{31}=\big\{2\big[(p_a^{+}p_a^{+}+q_a^{-}r_a^{-})m^{1/2}+2(8p_a^0p_b^0+p_a^{+}p_b^{+}+p_a^{-}p_b^{-})l^{1/2}\big](e^{i2\psi}q_b^{-}+e^{-i2\psi}r_b^{-})\big\}_{-}\\
\nonumber
&&+\big\{2\big[(p_a^{+}p_b^{-}+4p_a^{-}p_b^{+})m^{1/2}+p_b^{+}p_b^{-}l^{1/2}\big](e^{-i2\psi}q_a^{+}+e^{i2\psi}r_a^{+})\big\}_{-}
+\big\{\big[16p_a^0(2p_a^0 p_b^{+}-3p_b^0 p_a^{+})\\
\nonumber
&&
-6p_a^{+}(q_a^{+} q_b^{+}+r_a^{+} r_b^{+})+2p_b^{+}(4q_a^{+} r_a^{+}-q_a^{-} r_a^{-})-4p_a^{-}(p_a^{+}p_b^{-}-2p_a^{-}p_b^{+})+3p_a^{+}(q_a^{-}q_b^{-}+r_a^{-}r_b^{-}-p_a^{+}p_b^{+})
\\
&&+(e^{-i4\psi}/2)q_a^{-}(4p_a^{+}r_b^{-}-3p_b^{+}q_a^{-})+(e^{i4\psi}/2)r_a^{-}(4p_a^{+}q_b^{-}-3p_b^{+}r_a^{-})\big]m^{1/2}\big\}_{+}\\
&&+\big\{\big[(e^{-i4\psi}/2)q_a^{-}(2e^{-i2\psi}q_a^{-}r_b^{-}-8e^{i2\psi}r_a^{+}r_b^{+})+(e^{i4\psi}/2)r_a^{-}(2e^{i2\psi}q_b^{-}r_a^{-}-8e^{-i2\psi}q_a^{+}q_b^{+})\big]m^{1/2}\big\}_{-},
\end{eqnarray*}
\begin{eqnarray*}
\nonumber
 \kappa_{32}=\big\{p_a^{+}\big[(p_a^{+}p_a^{+}+2q_a^{-}r_a^{-})
 +(3/2)(e^{-i4\psi} q_a^{-}q_a^{-}+e^{i4\psi}r_a^{-}r_a^{-})\big]m^{3/2}\big\}_{+}
 +\big\{(2p_a^{+}p_a^{+}+q_a^{-}r_a^{-})(e^{-i2\psi} q_a^{-}+e^{i2\psi}r_a^{-})m^{3/2}\big\}_{-},\\
\end{eqnarray*}
\begin{eqnarray*}
\nonumber
 \kappa_{41}&=&32 \big[2 p_a^0 p_a^0(p_b^{-}-e^{i2\psi}q_b^{+})(p_b^{-}-e^{-i2\psi}r_b^{+})
   +2 p_b^0 p_b^0(p_a^{-}+e^{-i2\psi}q_a^{+})(p_a^{-}+e^{i2\psi}r_a^{+})\\
     \nonumber
   &+& p_a^0 p_b^0\big((p_a^{+}+e^{-i2\psi}q_a^{-})(p_b^{+}-e^{-i2\psi}r_b^{-})+(p_a^{+}+e^{i2\psi}r_a^{-})(p_b^{+}-e^{i2\psi}q_b^{-})\big)\big]\\
   \nonumber
   &+& 256 p_a^0 p_a^0p_b^0 p_b^0
   +(p_a^{+}+e^{-i2\psi}q_a^{-})^2(p_b^{+}-e^{-i2\psi}r_b^{-})^2
   +(p_a^{+}+e^{i2\psi}r_a^{-})^2(p_b^{+}-e^{i2\psi}q_b^{-})^2\\
    \nonumber
   &+& 4\big[(p_a^{+}+e^{-i2\psi}q_a^{-})(p_b^{+}-e^{i2\psi}q_b^{-})
-2(p_a^{-}+e^{-i2\psi}q_a^{+})(p_b^{-}-e^{i2\psi}q_b^{+})\big]\\
   &&\times\big[(p_a^{+}+e^{i2\psi}r_a^{-})(p_b^{+}-e^{-i2\psi}r_b^{-})
-2(p_a^{-}+e^{i2\psi}r_a^{+})(p_b^{-}-e^{-i2\psi}r_b^{+})\big],
\end{eqnarray*}
\begin{eqnarray*}
\nonumber
 \kappa_{42}=-32p_a^0p_b^0\big[(p_a^{+}+e^{-i2\psi}q_a^{-})(p_a^{+}+e^{i2\psi}r_a^{-})m
 +(p_b^{+}-e^{i2\psi}q_b^{-})
(p_b^{+}-e^{-i2\psi}r_b^{-}) l ]\hspace{2cm}\\
 \nonumber
- 2\big[(p_a^{+}+e^{-i2\psi}q_a^{-})(p_b^{+}-e^{i2\psi}q_b^{-})
-2(p_a^{-}+e^{-i2\psi}q_a^{+})(p_b^{-}-e^{i2\psi}q_b^{+})\big]
\big[(p_a^{+}+e^{i2\psi}r_a^{-})^2 m
 +(p_b^{+}-e^{-i2\psi}r_b^{-})^2 l\big]\\
 - 2\big[(p_a^{+}+e^{i2\psi}r_a^{-})(p_b^{+}-e^{-i2\psi}r_b^{-})
-2(p_a^{-}+e^{i2\psi}r_a^{+})(p_b^{-}-e^{-i2\psi}r_b^{+})\big]
\big[(p_a^{+}+e^{-i2\psi}q_a^{-})^2 m
 +(p_b^{+}-e^{i2\psi}q_b^{-})^2 l \big],
\end{eqnarray*}
\begin{eqnarray*}
 \kappa_{43}=(p_a^{+}+e^{-i2\psi}q_a^{-})^2(p_a^{+}+e^{i2\psi}r_a^{-})^2 m^2
 +(p_b^{+}-e^{i2\psi}q_b^{-})^2(p_b^{+}-e^{-i2\psi}r_b^{-})^2 l^2.
\end{eqnarray*}
In the above equations, we introduced the notation $\{\mathcal{E}(a,b,l,m,\psi)\}_{\pm}:=\mathcal{E}(a,b,l,m,\psi)\pm\mathcal{E}(b,a,m,l,-\psi)$ with $\mathcal{E}$ an expression involving $a,b,l,m,\psi$.

\end{document}